\documentclass[10pt, oneside]{article}   	
\usepackage{geometry}                		
\usepackage{graphicx}				
\usepackage{amssymb,amsmath} 
\usepackage{graphicx}
\usepackage[colorlinks=true,urlcolor=blue]{hyperref}

\usepackage{slashed}

\def\={\equiv}
 
\def\b{\beta} 

\def\c{\chi}
\def\d{\delta}

\def\f{\phi} 
 
\def\g{\gamma}
\def\h{\eta} 

\def\k{\kappa}

\def\l{\lambda} 
\def\m{\mu} 
\def\n{\nu}
\def\o{\omega} 
\def\p{\pi}

\def\vq{\vartheta} 
\def\r{\rho}
 
\def\s{{\sigma}} 

\def\t{\tau}

\def\y{\psi} 
\def\z{\zeta}

\def\D{\Delta} 
\def\F{\Phi} 
\def\G{\Gamma}
\def\L{\Lambda} 
\def\O{\Omega}

\def\S{\Sigma}

\newcommand{\0}[1]{{(#1)}}
\newcommand{\1}[1]{{\hat #1}}
\newcommand{\2}[1]{{\tilde #1}}
\newcommand{\3}[1]{{\boldsymbol #1}}

\newcommand{\7}[1]{{\bar#1}}

\newtheorem{defin}{Definition}
\newtheorem{prop}{Proposition}
\newtheorem{thm}{Theorem}

\newtheorem{rem}{Remark}

\newcommand{\HB}{\hfill\break}
\newcommand{\VE}{\vfill\eject}

\newcommand{\blue}{\color{blue}}

\newcommand{\lab}{\label}
\newcommand{\eq}{\eqref}
\newcommand{\ci}{\cite}
\newcommand{\bx}[1]{\boxed{\ #1\ }}

\newcommand{\bul}{$\bullet\ $}
\newcommand{\tr}{{\rm Tr\,}}

\newcommand{\app}{\approx} 
\newcommand{\rr}[1]{{{\mathbb R\hskip.5pt}^{#1}}}
\newcommand{\cc}[1]{{{\mathbb C\hskip.5pt}^{#1}}}

\newcommand{\pl}{\partial}

\newcommand{\w}{\wedge}

\newcommand{\dd}{{\rm d}}
\newcommand{\ie}{{\rm i.e., }}

\newcommand{\re}{{\,\rm Re}\ }   
\newcommand{\im}{{\,\rm Im}\ }  
\newcommand{\imp}{\ \Rightarrow\ }
\newcommand{\inv}{^{-1}}

\newcommand{\lra}{\leftrightarrow}

\newcommand{\Iff}{\Longleftrightarrow}
\newcommand{\mt}{{\,\mapsto\,}}
\newcommand{\pd}[2]{{\frac{\!\partial #1}{\partial#2}}}
\newcommand{\plra}{\,\pl^{\kern-1.25ex^\lra}}
\newcommand{\qq}{\quad} 
\newcommand{\qqq}{\qquad}

\newcommand{\sgn}{{\,\rm sgn \,}}
 
\newcommand{\sr}{\sqrt}
\newcommand{\stack}[2]{\stackrel{#1}{#2}} 

\newcommand{\sv}[1]{\vskip#1ex}
\newcommand{\Tr}{\mathrm{Tr\,}}

\newcommand{\lp}{\left(}
\newcommand{\rp}{ \right)}
\newcommand{\lb}{ \left[}
\newcommand{\rb}{\right]}
\newcommand{\la}{\langle\,}
\newcommand{\ra}{\,\rangle}

\newcommand{\LB}{\left\lbrace}
\newcommand{\RB}{\right\rbrace}

\newcommand{\bb}[1]{{\boldsymbol{\bar #1}}}
\newcommand{\bh}[1]{{\boldsymbol{\hat #1}}}
\newcommand{\bt}[1]{{\boldsymbol{\tilde #1}}}

\begin{document}

\title{T${\blue\hbar}$ermal Spacetime, Part I: \\
Relativistic Bohmian Mechanics}

\author{{Gerald Kaiser}\\
www.wavelets.com\\
kaiser@wavelets.com
}

\maketitle

\begin{abstract}
By complexifying Minkowski space $\4R^{1+d}$, the proper distance $\s\0x$ and proper time $\t\0x$ extend to the real and imaginary parts $\s\0z$ and $\t\0z$ of the \sl complex length \rm $\z\0z\=\sr{-z^2}$ of $z=x-iy$  (Fig.~\ref{F:Re-Im-zeta1}). For holomorphic positive-energy solutions of the Klein-Gordon equation to exist, $y$ must belong to the future cone $V_+$, thus forming a \sl local arrow of time \rm  without the need to invoke statistical physics.

The \sl future tube \rm $\5T_+=\4R^{1+d}-iV_+$ acts as an extended  phase space for the associated \sl classical \rm particle, the two extra variables being the time $x_0$ and $\l=\sr{y^2}>0$. The evaluation maps $e_z\colon \y\mt\y\0z$ on the space $\5K$ of holomorphic wave functions define a family of \sl fundamental states, \rm $e_z$ being the \sl quantization \rm  of $z\in\5T_+$ (Section \ref{S:quantization}) whose nonrelativistic limit is a Gaussian coherent state at time $x_0=0$ evolving relativistically to $x_0\ne 0$; see Figure \ref{F:Husimi}. A norm is defined in $\5K$ by $\|\y\|^2=\int_\G\dd\g\0z\,|\y\0z|^2$ where $\G$ and $\dd\g\0z$ are covariant forms of classical phase space and Liouville measure, respectively \eq{IPgen01}.  We prove that $\|\y\|^2$ is the total conserved charge of a \sl microlocal probability current \rm $j_\m\0z$, which implies that $\|\y\|$ is identical to the momentum space norm and $|\y\0z|^2$ is a probability density on \sl all \rm phase spaces $\G$ (covariant Born rule). This solves a long-standing problem in Klein-Gordon theory. The fundamental states $e_z$ give resolutions of unity for every $\G$ \eq{RU01}, generalizing those for the non-relativistic coherent states.  All of this generalizes to Dirac particles \ci[Chapter 5]{K90}.

A direct connection with thermal physics is established in Theorem \ref{T:zthermal}, where it is shown that the average of an operator $A$ in a relativistic canonical ensemble at the reciprocal temperature $\b$ in a reference frame with its time axis along the unit vector $u\in V_+$ is an integral of $\2A(z-i\vq)$ over $z\in\G$, where $\2A\0z=\la e_z|A| e_z\ra$, $\vq\=\tfrac12 \hbar\b u$ is the \sl thermal vector \rm specifying a quantum equilibrium frame and its temperature, and $\G$ is any covariant phase space. This proves that the \sl ensemble \rm of the thermal approach is the family of all  ``hidden'' phase-space trajectories of the associated classical particle. 

Interactions with gauge fields are included through \sl holomorphic gauge theory \rm (Section \ref{S:HGT}), which modifies the canonical ensemble by introducing a \sl fiber metric \rm $g\0z$ in the quantum Hilbert space.

\end{abstract}

\clearpage
\begin{center}
    \thispagestyle{empty}
    \vspace*{\fill}
    {\large\it
   For Angela,\\ \sv2
   With Love \& Gratitude
     }
    \vspace*{\fill}
\end{center}
\clearpage

\tableofcontents

\section{A Problem with Minkowski space}\label{S:prob}

Flat spacetime in $D=1+d$ dimensions is an \sl affine \rm space equivalent, as a set, to $\rr D$. It is not a vector space because no privileged event exists playing the role of origin. Rather, any two events $a,b$ can be connected by the vector $x$ called the \sl spacetime interval from $a$ to $b$, \rm which we write as a \sl row vector. \rm The set of all such intervals forms a vector space $M$ called \sl Minkowski space. \rm  We take the coordinates of $x$ to be $x_\m$ $(0\le \m\le d)$. Its time-space decomposition is
\begin{align}\lab{tx0}
x=(t,\3x)\ \ \text{where}\ \ 
t=x_0\in\4R\ \ \text{and}\ \ \3x=(x_1,\cdots, x_d)\in\rr d.
\end{align}
The \sl Minkowski scalar product \rm of two intervals $x,y\in M$ is given by
\begin{align}\lab{MSP}
x\cdot y&=x_0 y_0-\3x\cdot\3y=\sum_{\m,\n=0}^d x_\m \h^{\m\n} y_\n\=x_\m \h^{\m\n} y_\n,
\end{align}
where units have been chosen so that the universal speed of light $c=1$ and 
\begin{align}\lab{gmn00}
\h^{\m\n}=\text{diag}(1,-1,-1,\cdots, -1)
\end{align}
is the \sl Minkowski pseudo-metric \rm on $M$. (We shall reinsert $c$ in select formulas when it helps with physical interpretation.) The last identity in \eq{MSP} illustrates \sl Einstein's summation convention, \rm where identical superscripts and subscripts in each term are automatically summed over their range.

The \sl Lorentz group \rm $\5G$ is the set of all linear maps
\begin{align}\lab{LT000}
\L\colon M\to M,\ \ x\mt x\L
\end{align}
which preserve the scalar product  \eq{MSP}, \ie
\begin{align}\lab{LT00}
(x\L)\cdot (y\L)=x\cdot y\ \ \forall x,y\in M.
\end{align}

\rem\lab{R:LeftAction} \rm The $D\times D$ matrix $\L$ must act
\sl to the left \rm  on the row vector $x$. If a second Lorentz transformation $\L'$ is applied, the combined action is
\begin{align}\lab{xLL0}
(x\L)\L'=x(\L\L'),
\end{align}
so the order of mappings is from left to right, the same as mathematical writing. The convention \eq{LT000} could thus be called \sl chronological. \rm Had we taken $x$ to be a \sl column vector, \rm the order of mappings would be \sl anti-chronological:\rm
\begin{align}\lab{xLL1}
\L'(\L x)=(\L'\L) x.
\end{align}
This explains our unconventional preference for row vectors and left-acting operators. The same will apply to quantum wave functions and operators.\ \ \rm $\clubsuit$

However, $\5G$ includes \sl space and time inversions. \rm Unless stated otherwise, we confine ourselves to the \sl restricted Lorentz group, \rm which excludes all inversions: 
\begin{align}\lab{5L0}
\5G_0=\{\L\in\5G\colon \det\L=1\  \text{and $\sgn(x\L)_0=\sgn x_0\ \forall x\in M$}\}.
\end{align}
The condition $\det\L=1$ ensures that the overall orientation of $M$ remains unchanged, while the invariance of $\sgn x_0$ ensures that the order of time is preserved, hence so is the orientation of space (since  $\det\L=1$). 

The \sl Minkowski quadratic form \rm is the mapping 
$Q\colon M\to\4R$ defined by
\begin{align}\lab{QMx00}
Q\0x=t^2-r^2\=x^2,\ \ \text{where}\ \ \ r=\sr{\3x\cdot\3x}=\sr{\3x^2}\ge 0.
\end{align}
Since $Q$ is indefinite, $M$ breaks into the three Lorentz-invariant sectors
\begin{equation}\lab{TLS0}\begin{split}
\text{Timelike intervals:}\ \ V&=\{x\in M\colon x^2>0\}\\
\text{Lightlike intervals:}\ \ L&=\{x\in M\colon x^2=0\}\\
\text{Spacelike intervals:}\ \ S&=\{x\in M\colon x^2<0\}
\end{split}\end{equation}
and $M$ is their disjoint union
\begin{align}\lab{MTLS0}
M=V\cup L\cup S.
\end{align}
$V$ and $L$ further break into the disjoint unions
\begin{equation}\lab{TLpm}\begin{split}
&V=V_+\cup V_- \qqq  V_\pm=\{(t,\3r)\colon \pm t>r\}\\
&L=L_+\cup L_- \qqq  L_\pm=\{(t,\3r)\colon \pm t=r\}
\end{split}\end{equation}
where
\begin{equation}\lab{Tpm00}\begin{split}
&V_+\ \ \text{is the \sl future cone}\ \ \qqq\qq\ 
V_-\ \ \text{is the \sl past cone}\\
&L_+\ \ \text{is the \sl future light cone}\ \ \qq
L_-\ \ \text{is the \sl past light cone}.
\end{split}\end{equation}
The physical significance of the decomposition \eq{MTLS0} is as follows.
\begin{enumerate}

\item Two distinct events $\{a,b\}$ can be made \sl simultaneous \rm by a Lorentz transformation if and only if their interval $x=b-a$ is spacelike:
\begin{align}\lab{simult}
x^2<0 \Iff x\L=(0,\3\s)\ \text{for some} \ \L\in\5G_0,\ \30\ne\3\s\in\rr d
\end{align}
and \eq{LT00}  then gives $x^2=-\3\s^2$. While $\3\s$ is not Lorentz invariant, its lenght is:
\begin{align}\lab{propdist}
|\3\s|=\sr{-x^2}\=\s\0x> 0,\ \ x\in S.
\end{align}
$\s\0x$ is then the \sl proper distance \rm between the events. 

\item Two distinct events  can be Lorentz-transformed to the same \sl spatial  \rm location if and only if their interval $x$ is timelike:
\begin{align*}
x^2>0\Iff x\L=(\t\0x,\30)\ \text{for some}\  \L\in\5G_0,\ \t\0x\ne 0,
\end{align*}
and \eq{LT00} then gives $x^2=\t\0x^2$ or
\begin{align}\lab{proptime0}
\t\0x=\pm\sr{x^2},\ x\in V.
\end{align}
While $|\t\0x|$ is the usual proper time interval between the events,  its sign identifies their \sl chronological order \rm if we set $\sgn \t\0x=\sgn t$. This leads to the following definition of \sl chronological proper time interval \rm between $a$ and $b$:
\begin{align}\lab{proptime}
\t\0x=\1t\sr{x^2},\ \ x\in V
\end{align}
where\footnote{The notation $\1t=t/|t|$ is just a one-dimensional version of the vector notation $\bh r=\3r/|\3r|$.
}
\begin{align}\lab{1t}
\1t\=\frac t{|t|}=\sgn t,\ (t^2>r^2\ge 0)
\end{align}
is invariant under all $\L\in\5G_0$.  $\t\0x$ is the chronologically oriented time interval between the events as measured by a clock whose (straight) worldline passes through both in the \sl future \rm direction. Since $x^2$ does not determine $\1t$, neither does it determine $\t\0x$. 

\item Any two events can be connected by a \sl light ray \rm if and only if their interval $x$ is lightlike:
\begin{align}\lab{xL0}
x^2=0 \Iff x\L=(\pm |\3\s|,\3\s)\ \text{for some} \ \L\in\5G_0,\ \3\s\in\rr d.
\end{align}
\end{enumerate}

\section{Thermal Spacetime and its Complex Length}\label{S:MPD}

Does a single function exist that is defined on all of $M$ and unifies the proper distance $\s\0x$ and proper time $\t\0x$? We shall see that it does --- but only if we are willing to give up \sl time reversal invariance \rm and allow our spacetime to include all possible \sl arrows of time. \rm The plural \sl arrows \rm is required by Relativity since all future-pointing arrows are equivalent under $\5G_0$, as are all past-pointing arrows. 

 An obvious starting point is the observation that
\begin{align}\lab{zeta000}
\sr{-x^2}=\begin{cases}
\s\0x> 0, & x^2< 0\\
\pm i\t\0x, & x^2>0
\end{cases}
\end{align}
where the sign in the timelike case is indeterminate since $x^2$ does not distinguish between past and future. We shall make sense of \eq{zeta000} by \sl complexifying \rm $x$ to
\begin{align}\lab{2x}
z=x-iy \ \ \text{with}\ \  y^2>0.
\end{align}
We call the set $\5T=M-iV$ of all such complex intervals the \sl causal tube. \rm 

\rem\lab{R:sectors}\rm
The causal tube is the disjoint union
\begin{align}\lab{5Tpm000}
\5T=\5T_+\cup\5T_-
\end{align}
where
\begin{equation}\lab{Tpm3}\begin{split}
\5T_+&=\{x-iy\in\cc D\colon x\in M, \ y\in V_+\}\ \ \text{is the \sl Future Tube}\\
\5T_-&=\{x-iy\in\cc D\colon x\in M, \ y\in V_-\}\ \ \text{is the \sl Past Tube}.
\end{split}\end{equation}
$\5T_+$ and $\5T_-$  play a central role in  quantum field theory \ci{SW64}, where they are called the \sl forward and backward tubes. \rm However, no attempt is made there to interpret $\5T_\pm$ physically, as will be done here; see also \ci{K90} and \ci{K11}.
\rm $\clubsuit$

\newcommand{\sebo}[1]{{\fontseries{sb}#1}}
\defin\lab{D:zeta0}
The {\bf complex length} of $z\in\5T$ is the analytic continuation of $\s\0x=\sr{-x^2}$ \eq{propdist} from $S\subset M$ to $\5T$  given by
\begin{align}\lab{zeta001}
\bx{\z(z)=\sr{-z^2}=\sr{y^2-x^2+2iy\cdot x}.}
\end{align}
The  {\bf extended proper distance} and {\bf extended chronological proper time} in $\5T$ are
\begin{align}\lab{st0001}
\s\0z=\re\z\0z \ \ \text{\rm and}\ \  \t\0z=\im\z\0z.
\end{align}
\rm

\rem\lab{R:Principal} \rm
Note that $\z\0z$ cannot vanish in $\5T$ since
\begin{align*}
\z\0z=0\imp y^2=x^2\ \ \text{and}\ \ y\cdot x=0,
\end{align*}
which is impossible since $x$, like $y$, is timelike.
Furthermore, 
\begin{align*}
y\cdot x=0\imp x^2<0\imp -z^2=y^2-x^2>0,
\end{align*}
hence $-z^2$ belongs to the the right-hand plane $\4C_+$ and $\z\0z$ belongs to the \sl cut \rm plane 
\begin{align*}
\4C_*=\sr{\4C_+}=\4C-N
\end{align*}
where $N$ is the negative real axis. In other words, $\z$ is the \sl principal branch \rm of $\sr{-z^2}$. \ $\clubsuit$

\rem\lab{R:Kz} \rm The most important role $\z\0z$ plays is in quantum theory, where it provides a measure of the distance between fundamental quantum states; see Eq.~\eq{SUM00}. \rm \ $\clubsuit$

\rem\lab{R:QL0}\rm
In Theorem \ref{T:zthermal}  we relate the new variable $y$ to the  \sl thermal vector \rm
\begin{align}\lab{vq00}
\vq=\tfrac12 \hbar\b u
\end{align}
where $\b=\sr{y^2}/\hbar$ is a relativistic analogue of the reciprocal equilibrium temperature in a quantum canonical ensemble and $u$ is the  $D$-velocity of the  associated equilibrium frame. This will be the basis of the \sl thermal spacetime \rm interpretation of $\5T$.
\ $\clubsuit$

\rem\lab{R:QuantEquil} \rm 
The above notion of  ``equilibrium'' for a single relativistic quantum particle is based on the fact that in our formalism, thermal expectations of operators can be represented as ensemble averages where the ensemble is simply the set of all relativistic phase-space trajectories of the associated \sl classical \rm particle (Theorem \ref{T:zthermal}). These are `hidden variables'  according to the Copenhagen interpretation; see Remark \ref{R:Ensemble}. This may be related to the notion of \sl quantum equilibrium \rm in Bohmian Mechanics and its connection to Born's rule \ci{DGZ92}; see  Remark \ref{R:RelBorn}. \rm $\clubsuit$

\prop\lab{stH0} 
The boundary value of $\z$ as $y\to 0$ in $\5T_\pm$  is the distribution 
\begin{align}\lab{Bdyvals0}
\lim_{y\to 0}\z(x-iy)=\s\0x H(-x^2)\pm i\t\0x H(x^2)
\end{align}
where $H$ is the Heaviside step function. This resolves the sign ambiguity in \eq{zeta000}.
\rm

\bf Proof: \rm
If $y\in V_\pm$ and $x^2<0$, then 
\begin{align*}
\lim_{y\to 0}\z(x-iy)=\lim_{y\to 0}\sr{y^2-x^2+2iy\cdot x}
=\sr{-x^2}=\s\0x.
\end{align*}
If $x^2>0$, let $\l=\sr{y^2}$ and use the invariance of $\z$ under $\5G_0$ to transform to a \sl rest frame \rm
\begin{align}\lab{RF0}
y=(\pm\l,\30)\in V_\pm\ \ \text{and}\  \ y\cdot x=\pm\l t.
\end{align}
Then
\begin{align}\lab{RF0}
\lim_{y\to 0}\z(x-iy)=\lim_{\l\to 0}\sr{\l^2-x^2\pm 2i\l t},
=\pm  i\1t\sr{x^2}=\pm i\t\0x.\ \blacksquare
\end{align}

\begin{figure}[h]\begin{center}
{\centering\includegraphics[width=4 in]{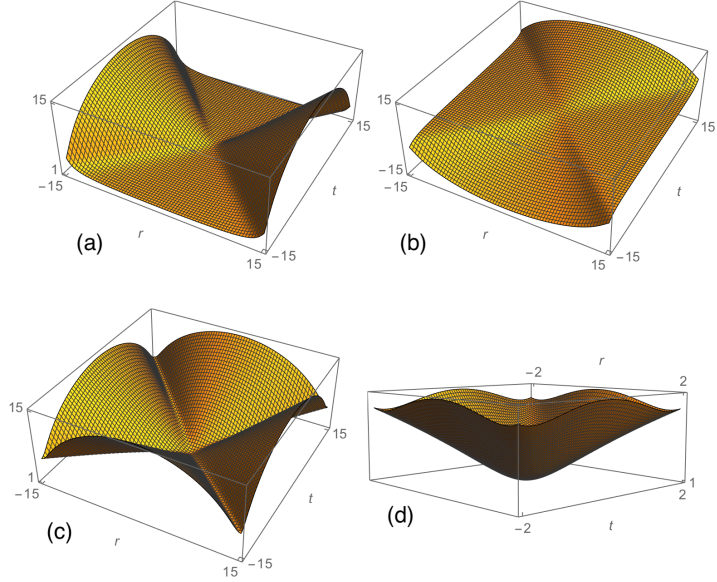}} 
\caption{Plots of $\s\0z$ and $\t\0z$ with $y=(1,\30)$: (a) $\s(t,r)$, (b) $\t(t,r)$, (c) $|\z(t,r)|$. Figure (d) is a closeup of $|\z(t,r)|$ showing the smooth minimum $\z(0,0)=1$. }\label{F:Re-Im-zeta1} \end{center}\end{figure}

Figure \ref{F:Re-Im-zeta1} shows plots of $\s,\t$, and $ |\z|$ with $D=2, y=(1,0)$ and $x=(t,r)\in\rr 2$, so that
\begin{align}\lab{zrt00}
\z(t,r)=\sr{1+r^2-t^2+2i t}\=\s(t,r)+i\t(t,r).
\end{align}

The level surfaces of $\s\0x$ in $S$ and $\t\0x$ in $V$  are the hyperboloids
\begin{equation}\lab{level0}\begin{split}
B_\s &=\{x\in S\colon r^2-t^2=\s^2\}\subset S,\ \s> 0\\
W_\t &=\{x\in V\colon t^2-r^2=\t^2\}\subset V,\ \t\ne 0.
\end{split}\end{equation}

\prop\lab{P:LevelSurf}
The level surfaces of $\s\0z$ and $\t\0z$ with $y=(\l,\30)$ are the hyperboloids
\begin{equation}\lab{BW00}\begin{split}
\5B_\s&=\LB x\in M\colon \frac{r^2}{\s^2-\l^2}-\frac{t^2}{\s^2}=1 \RB\ \ \s>0\\
\5W_\t&=\LB x\in M\colon \frac{t^2}{\t^2}-\frac{r^2}{\t^2+\l^2}=1,\ \1t=\1\t\RB,\ \ \t\ne 0
\end{split}\end{equation}
where the condition $\1t=\1\t$ eliminates the chronologically dissonant half of the  two-sheeted hyperboloid.  The intersection
\begin{align}\lab{Xst}
X_{\s,\t}\=\5B_\s\cap\5W_\t
\end{align}
is the level set of the complex distance:
\begin{align}\lab{zst0}
X_{\s,\t}=\{z\in\5T\colon \z\0z=\s+i\t\}.
\end{align}

\rm

As expected,
\begin{align}\lab{lambdazero}
\l\to 0\imp \5B_\s\to B_\s\ \text{and}\ \5T_\t\to V_\t.
\end{align}
\rm

\bf Proof: \rm
For $z\in\5T_+$, we can choose $y=(\l,\30)$. Then
\begin{align*}
\z=\sr{\l^2-t^2+r^2+2i\l t}=\s+i\t,
\end{align*}
from which
\begin{align*}
\l^2-t^2+r^2=\s^2-\t^2\ \ \text{and}\ \ \l t=\s\t
\end{align*}
hence
\begin{align*}
\l^2r^2&=\l^2(t^2+\s^2-\t^2-\l^2)\\
&=\s^2\t^2+\l^2\s^2-\l^2\t^2-\l^4.
\end{align*}
The right side factorizes, giving 
\begin{align}\lab{sigtauTR}
\l^2r^2=(\s^2-\l^2)(\t^2+\l^2)\qqq  \l t=\s\t.
\end{align}
This proves that $(\s,\t)$ carries information equivalent to $(t,r)$ in $\5T_+$. Hence
\begin{align*}
\frac{r^2}{\s^2-\l^2}=\frac{\t^2}{\l^2}+1\qqq
\frac{r^2}{\t^2+\l^2}=\frac{\s^2}{\l^2}-1,
\end{align*}
and \eq{BW00} follows from $\l t=\s\t$. \ $\blacksquare$

Equations \eq{BW00} and \eq{sigtauTR} are not Lorentz-invariant because we chose $y=(\l,\30)$ from the outset. This is easily remedied.

\defin\lab{D:LocalTR} 
Given $y\in V$, let
\begin{align}\lab{LocalTR0}
\l=|y|\=\sr{y^2}\ \ \text{and}\ \ \1y=y/\l.
\end{align}
The invariant local time and radial coordinates relative to $y$ are
\begin{align}\lab{LocalTR1}
t_y\0x=\1y\cdot x\ \ \text{and}\ \ r_y\0x=\sr{t_y^2-x^2}.
\end{align}
\rm

Note that
\begin{align*}
t_y^2-r_y^2=x^2=t^2-r^2
\end{align*}
and
\begin{align*}
y\to (\pm\l,\30)\in V_\pm \imp \{t_y\0x\to\pm t,\ \ r_y\0x\to r\}.
\end{align*}
By choosing any $y\in V_+$ and substituting $t_y$ for $t$ and $r_y$ for $r$, Equations \eq{sigtauTR} take the invariant form
\begin{align}\lab{sigtauTR2}
\l^2r_y^2=(\s^2-\l^2)(\t^2+\l^2)\qqq  \l t_y=\s\t
\end{align}
relating the \sl local \rm invariants $(t_y\0x, r_y\0x)$ to the \sl global \rm invariants $(\s\0z,\t\0z)$.

\rem\lab{R:K00} \rm A different route to complex spacetime was developed in \ci{K00}. \  $\clubsuit$

\section{The Quantization of $\5T_+$}\lab{S:quantization}

\begin{quote}
\sl We conclude that the Klein-Gordon equation does not have a consistent single-particle interpretation and the naive transcription of the trajectory interpretation of nonrelativistic Schrödinger quantum mechanics into this context does not work. \rm \ -- Peter Holland in \ci{H93}.
\end{quote}

Here we resolve this well-known problem by quantizing a Klein-Gordon particle in the future tube $\5T_+$, interpreted as an extended phase space. In the process we discover that the quantum randomness in this case is due to averaging ``observables'' over a \sl hidden\,\rm\footnote{The ensemble is ``hidden'' because classical trajectories are not an admissible quantum concept.
} 
canonical ensemble consisting of all classical phase-space particle trajectories. \rm  This amounts to a phase-space formulation of \sl relativistic Bohmian Mechanics. \rm 

\rem\lab{R:Dirac} \sl Simplified Dirac notation. \rm 
Let $\6H$ be a complex Hilbert space with inner product $\la f | g\ra$ linear in $f$ and antilinear in $g$.\footnote{This convention works well with the left action of operators. Physicists use the opposite convention.
}
If $\6H$ were finite-dimensional, then the inner product of the row vectors $f,g$ could be expressed in matrix form as 
\begin{align}\lab{DN3}
\la f | g\ra=fg^\dag
\end{align}
where the column vector $g^\dag$ is the Hermitian conjugate of $g$. This can be extended to infinite dimensions in a mathematically rigorous way \ci{K11}. We adapt $fg^\dag$ as a simplified form of Dirac's \sl bra-ket \rm notation $\la f | g\ra$.  \ $\clubsuit$

A \sl massive scalar \rm is a single free spinless relativistic particle of mass $m>0$.  A  \sl plane wave \rm with energy-momentum $p=(E,\3p)$ is given by
\begin{align}\lab{PW0}
\f_p\0x=e^{-ix\cdot p/\hbar}
=e^{(itE-i\3x\cdot\3p)/\hbar},\qq E=\sr{m^2+\3p^2}.
\end{align}
This is the beginning of quantum mechanics. It associates with a particle of energy-momentum $p$ a wave of frequency  $k_0$ and wave vector $\3k$ given by the \sl Planck–Einstein-de Broglie relations \rm
\begin{align}\lab{pk1}
k_0=E/\hbar,\qq \3k=\3p/\hbar.
\end{align} 
Since the particle is free, $p$ belongs to the \sl mass shell \rm
\begin{align}\lab{Om0}
\O_m=\{(E,\3p)\colon E=\sr{m^2+\3p^2},\ \3p\in\rr d\}.
\end{align}
Since all $p\in\O_m$ must have equal weight by Einstein's Relativity Principle and $\3p$ varies over $\rr d$, each $p$ has weight zero. This means that $\O_m$ must be treated as a measure space, where the `weight' of a measurable subset $A\subset\O_m$ is its measure
\begin{align}\lab{WA0}
\m(A)=\int_A\dd\m\0p.
\end{align}
For $\m\0A$ to be frame-independent, $\dd\m$ must be Lorentz-invariant.  To find it, note that for general $p=(p_0,\3p)\in V_+$ we have $p^2-m^2=p_0^2-E^2$, hence
\begin{align*}
\d(p^2-m^2)\dd p &=\d((p_0-E)(p_0+E))\dd p_0\,\dd\3p
=\d(p_0-E)\,\dd p_0\,\frac{\dd\3p}{2E}\,,
\end{align*}
proving that $\dd\m$ is given uniquely, up to a constant factor, by
\begin{align}\lab{dmu0}
\dd\m\0p=\frac{\dd\3p}{2E}\qq (p\in\O_m).
\end{align}
The numerator $\dd\3p$ is the Galilean-invariant Lebesgue measure on the nonrelativistic momentum space $\rr d$, and the denominator $2E$ accounts for the \sl curvature \rm of the hyperboloid $\O_m$. Momenta $\3p$ with large energies $E(\3p)$ count for less in $\dd\m$ than they would in $\dd\3p$, thus making the space of integrable functions larger:
\begin{align}\lab{L10}
L^1(\dd\m)\supset L^1(\dd\3p).
\end{align}

\rem\lab{R:L2Rel}\rm
The curvature factor $(2E)\inv$ in $\dd\m\0p$ breaks the 
symmetry between the position and momentum representations of nonrelativistic quantum mechanics, on which the canonical commutation relations and the Heisenberg Uncertainty Principle are based. That complicates many aspects of the theory, including the inner product in the position representation (as compared with the momentum representation \eq{L2fp2}, which is straightforward), the spatial probability interpretation, and even the existence of position operators. This results in the well-known non-existence of a covariant probability interpretation for massive scalar particles in \sl real \rm spacetime, which will be resolved in thermal spacetime; see also \ci[Chapter 4]{K90}. \rm \ $\clubsuit$

The plane wave $\f_p\0x$ extends to the entire function
\begin{align}\lab{PW2}
\f_p\0z=e^{-iz\cdot p/\hbar}=\f_p\0x e^{-y\cdot p/\hbar},\ \ z=x-iy\in\cc D
\end{align}
satisfying the holomorphic Klein-Gordon equation
\begin{align}\lab{HolKG0}
-\Box _z\f_p\0z
\=-\frac{\pl^2\f_p\0z}{\pl z^\m\pl z_\m}=(mc/\hbar)^2\f_p\0z.
\end{align}
But what happens to a general superposition of such plane waves? The question about the compatibility of the complexification 
$M\to\5T_+$ with quantum theory thus comes down to studying the behavior of the function
\begin{align}\lab{Wy00}
R_y\colon \O_m\to\4R,\qq R_y\0p=e^{-y\cdot p/\hbar},
\ \text{where}\ y\in\rr D\ \text{and}\ p\in\O_m.
\end{align}
A general holomorphic solution of \eq{HolKG0} with positive energy is a continuous superposition of holomorphic plane waves $\f_p\0z$ with all possible $p\in\O_m$, 
\begin{align}\lab{phiz00}
\y\0z=\int_{\O_m}\dd\m\0p\,a\0p \f_p\0z.
\end{align}

We call $\y\0z$ and $a\0p$ the \sl $z$-representation \rm and \sl $p$-representation \rm of the quantum state, respectively.
The Hilbert space of the $p$-representation is
\begin{align}\lab{L2fp0}
\5H\=L^2(\dd\m)=\LB a\colon \O_m\to\4C, \ \| a\|<\infty\RB
\end{align} 
where the norm $\|a\|\ge 0$ is given by
\begin{align}\lab{L2fp1}
\| a\|^2\=\int_{\O_m}\dd\m\0p\,|a\0p|^2
\end{align}
with inner product \eq{DN3}
\begin{align}\lab{L2fp2}
a_1a_2^\dag\=\int_{\O_m}\dd\m\0p\,a_1\0p a_2\0p^*.
\end{align}
The Hilbert space of the $z$-representation is 
\begin{align}\lab{5K00}
\5K=\{\y\0z=\int_{\O_m}\!\!\dd\m\0p\, a\0p e^{-iz\cdot p/\hbar}\colon a\in \5H\}
\end{align}
with inner product imported, initially, from $\5H$:
\begin{align}\label{IP000}
\y_1\y_2^\dag\equiv a_1 a_2^\dag.
\end{align}
Clearly, $z$ must be confined to $\5T_+$  for $\y\0z$ to converge when $a\in\5H$. In that case, $\y\0z$ is a \sl holomorphic positive-energy solution \rm of the holomorphic Klein-Gordon equation
\begin{align}\lab{HKG0}
-\Box _z\y\0z=(mc/\hbar)^2\y\0z.
\end{align}
We shall express $ \y_1\y_2^\dag $ as an integral over a relativistic classical phase space  $\G\subset\5T_+$ of  dimension 
$\dim_{\4R}\G=2d$. This will give a Lorentz-covariant probability interpretation of $\y\0z$ generalizing the Born rule. As noted before, such an interpretation is missing \rm in $M$. 

To see how $\y\0z$ and $a\0p$ transform under the restricted Lorentz group $\5G_0$,\footnote{Here we must confine ourselves to the \sl reduced \rm Lorentz group $\5G_0$ in order to leave $\5T_\pm$ invariant, as  is necessary by Proposition \ref{P:scalar}.
}
we must first explain how $z$ transforms. The action of $\5G$ on $M$ extends to $\cc D$ by \sl complex linearity, \rm \ie
\begin{align}\lab{zL00}
z\L\=x\L -i y\L,\qq z=x-iy\in\cc D,\qq \L\in\5G.
\end{align}
Since $\5T_+$ is not invariant under $z\mt -z$, we must confine our analysis to the  restricted Lorentz group $\5G_0$, whose actions on $a\0p$ and $\y\0z$ are given by
\begin{align}\lab{Lf00}
aU(\L)\0p=a(p\L)\qq \text{and}\qq \y U(\L)\0z=\y(z\L),
\end{align}
from which
\begin{align}\lab{U1U20}
U(\L_1\L_2)=U(\L_1)U(\L_2)
\end{align}
as required of a representation. From the invariance of $\O_m$ and $\dd\m$ it follows that the $p$-representation is unitary, hence so is the $z$-representation by \eq{IP000}.

\prop\lab{P:RTI} Reverse Triangle Inequality {\rm (Figure \eq{F:RayFilters2}).}

If $y$ and $p$ are any vectors in $V_+$, then
\begin{align}\lab{RTI0}
y\cdot p\ge |y| |p|\ \ \text{where}\ \ |y|=\sr{y^2} \ \ \text{and}\ \  |p|=\sr{p^2},
\end{align}
with equality if and only if $y$ and $p$ are parallel:
\begin{align}\lab{py00}
y\cdot p= |y| |p| \Iff \1p=\1y
\end{align}
where $\1p=p/|p|$ and $\1y=y/|y|$.
\rm

\bf Proof: \rm Choose a `rest frame' with $y=(|y|,\30)$. Then
\begin{align}\lab{py01}
y\cdot p=|y|\sr{|p|^2+\3p^2}\ge |y| |p|
\end{align}
with equality if and only if $\3p=\30$, in which case
\begin{align*}
p=(|p|,\30)=|p|\1y\imp\1y=\1p.
\end{align*}
By the $\5G_0$-invariance of $y\cdot p$, this is true in any inertial frame.\  $\blacksquare $

\sv{-10}
\begin{figure}[h]\begin{center}
{\centering\includegraphics[width=3 in]{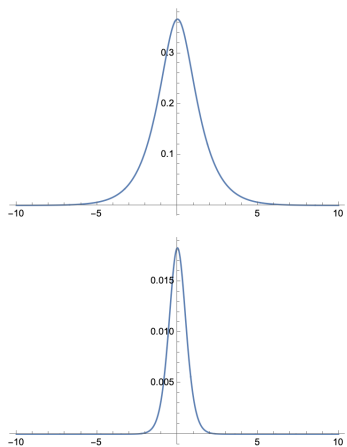}}
\sv{-14}
\caption{The ray filter \eq{Slambda0} in $d=1$ space dimension with  $m=\hbar=1,\ y=(\l,0)$, and $p=(\sr{1+q^2},q)$, thus $S_\l\0q=e^{-\l\sr{1+q^2}}$ in \eq{Slambda0}. The upper and lower plots show  $S_1\0q$ and $S_4\0q$, demonstrating the increasing \sl directivity \rm of $R_y$ with $\l$.}
\label{F:RayFilters2}\end{center} \end{figure}

\rem\lab{R:WFT} $\y\0z$ as a Relativistic Windowed Fourier Transform. 

A \sl Windowed Fourier Transform \rm of $f\colon \rr n\to\4C$ has the form 
\begin{align}\lab{WFT0}
\2f(\3x,\3q)=\int_{\rr n}\dd\3k\,\1f(\3k)e^{i\3x\cdot\3k} W(\3k-\3q),
\end{align}
where $\1f(\3k)$ is the Fourier transform of $f(\3x)$ and $W(\3k)$ is a \sl window \rm centered around the origin.
The translates $W(\3k-\3q)$ of $W(\3k)$ \sl filter \rm$\1f$ down to a neighborhood of $\3q$ before applying the inverse transform.\footnote{The roles of $\3k$ and $\3x$ can also be interchanged, in which case a \sl spatial \rm window $W(\3x-\3y)$ reduces $f(\3x)$ to a neighborhood of $\3y$ before computing the Fourier transform. However, \eq{WFT0} is the correct choice in the case \eq{phiz00} since $a\0p$ is in the Fourier domain. See [K11] for a detailed exposition of windowed Fourier transforms, frames, and related matters.
}
Let us compare \eq{WFT0} with \eq{phiz00}, written in the form  
\begin{align}\lab{phiz01}
\y(x,y)=\int_{\O_m}\dd\m\0p\,a\0p e^{-ix\cdot p/\hbar} R_y\0p.
\end{align}
If $y$ is restricted to a single hyperboloid $\O_\l\subset\5T_+$ \eq{Olambda0}, then for any $y,y'\in\O_\l$ there exists $\L\in\5G_0$ such that $y'=y\L$ and any two windows are related by a Lorentz transformation:
\begin{align}\lab{Ry00}
R_{y'}\0p=R_{y\L}\0p=R_y(p\L\inv).
\end{align}
Thus \eq{phiz01} may be called a \sl Relativistic Windowed Fourier Transform. \rm By comparison, since any two  windows $W(\3k-\3q)$ and $W(\3k-\3q')$ in \eq{WFT0} are related by a translation, \eq{WFT0} may be called a \sl Euclidean \rm Windowed Fourier Transform. \ \ \rm $\clubsuit$

\prop\lab{P:scalar}
For a free massive scalar, the following are true:
\begin{enumerate}
\item $\y\0z$ is holomorphic for all $a\in \5H$ if and only if $z\in\5T$ is restricted to  $\5T_+$.
\item $R_y\0p$ filters $a\0p$ down to a {\bf ray bundle} centered around the direction $\1p=\1y$. 
\item $R_y\0p$ is a `bump function' on $\O_m$ peaking at 
\begin{align}\lab{py10}
p_y=m\1y.
\end{align}
\item $R_y\0p$ is a {\bf guiding filter} for the wave $\y(x-iy)$, steering it along $\1y\in V_+$.
\item $\l\=\sr{y^2}$ is a measure of the {\bf directivity} of $R_y$: the greater $\l$, the more narrowly the filter is focused around its maximizing direction $\1y$. 
\end{enumerate}
Thus, all $D$ components of $y$ are physically significant. 
\rm

\bf Proof: \rm If $y\in V_\pm$, choose a `rest frame' where $y=(\pm\l,\30)$. Then
\begin{align}\lab{Slambda0}
R_y\0p=e^{\mp \l \sr{m^2+\3p^2}}\= S_{\pm\l}(\3p)
\end{align}
where we have set $\hbar=1$ for convenience.
$S_{-\l}(\3p)$ grows as $e^{\l |\3p|}$,  ruling out $y\in V_-$. For $y\in V_+$, $S_\l$ decays as $e^{-\l |\3p|}$ and the integral \eq{phiz00} converges absolutely for all $a\in \5H$, defining the function $\y\0z$. It remains absolutely convergent when differentiated with respect to $z^\m$ under the integral sign, so $\y$ is holomorphic in $\5T_+$.
To prove the other points, choose $y=(\l,\30)\in V_+$. Then \eq{Slambda0} becomes
\begin{align*}
R_y\0p=e^{-\l\sr{m^2+\3p^2}}\=S_\l(\3p),
\end{align*}
which guides the wave function along a ray bundle centered about $\1p=\1y$.  The filter $S_\l(\3p)$ becomes exponentially sharper with increasing $\l$, as seen in Figure \eq{F:RayFilters2}.
Again, the above proofs are independent of the choice $y=(\l,\30)$  due to $\5G_0$-invariance.
\ $\blacksquare$

\rem\lab{R:Bohmian} \rm
Proposition \ref{P:scalar} suggests a connection to the de Broglie–Bohm pilot wave theory,  \ci{DGZ92,H93}, but with a fundamental difference:

 \sl The pilot is built into the underlying geometry through $y\in V_+$ and its guiding property follows from the holomorphy of $\y(x-iy)$.  \rm
 
Our theory so far is restricted to a single free relativistic particle. The next steps are to 
\begin{enumerate}
\item extend the theory to $N$ identical and independent free particles;
\item extend further to $N=\infty$ and relate this to a free quantum field theory;
\item find a way to include gauge interactions without destroying  holomorphy.
\end{enumerate}
These tasks should be guided by the fact that $\5T_+$ is the basis for axiomatic as well as constructive quantum field theory \ci{SW64, GJ87}.  \   $\clubsuit$

\rem\lab{R:Locality} \rm While $\y\0z$ is a solution of the Klein-Gordon equation \eq{HKG0}, this is not the whole story because it does not explicitly state that it is a \sl positive-energy \rm solution.  That fact can be included by requiring that $\y\0z$ be a solution of the psuedo-differential equation
\begin{equation}\lab{PosEn00}\begin{split}
\sr{-\Box_z}\,\y\0z=(mc/\hbar)\y\0z,
\end{split}\end{equation}
which is \sl non-local. \rm However, locality can be restored in $\5T_+$  if we replace the  \sl positive energy \rm requirement with \sl holomorphy, \rm expressed by the Cauchy-Riemann equations 
\begin{align}\lab{CR0}
\7\pl_\m\y\0z\=\pd{\y\0z}{\7z^\m}=0.
\end{align}
Then $\y\0z$ is simultaneously a solution of the equations \eq{HolKG0} and \eq{CR0} in $\5T_+$, both of which are \sl local in $\5T_+$. \rm Note that the equations remain non-local in $M$.  $\clubsuit$

\subsection{Review of Nonrelativistic (Gaussian) Coherent States}\lab{SS:GCS}

We shall see that the $z$-representation is closely related to nonrelativistic coherent-states representations, which will now be reviewed.

Consider a nonrelativistic particle in $\rr d$, whose position and momentum operators $\3X,\3P$ satisfy the canonical commutation relations
\begin{align}\lab{NRCS0}
[X_j, X_k]=[P_j, P_k]=0,\qq [X_j, P_k]=i\hbar\d_{jk},\qq 1\le j,k\le d
\end{align}
and act on a wave function $f(\3x)\in L^2(\rr d)$ and its Fourier transform $\1f(\3p)$ by
\begin{equation}\lab{NRCS00}\begin{split}
fX_k(\3x)&=x_k f(\3x)\qqq\  fP_k(\3x)=-i\hbar\pd{f(\3x)}{x_k}\\
\1fX_k(\3p)&=i\hbar\pd{\1f(\3p)}{p_k}\qqq \1f P_k(\3p)=p_k \1f(\3p).
\end{split}\end{equation}
To construct coherent states, fix any real number $\k$ and let
\begin{align}\lab{NRCS1}
A_k=X_k+i\k P_k.
\end{align}
Given a normalized state $f$, define $z\in\cc d$ by
\begin{align}\lab{NRCS2}
\7z_k\=f A_k f^\dag=\la A_k\ra_{\! f}
=\la X_k\ra_{\! f}+i\k \la P_k\ra_{\! f}=\2x_k+i\k \2p_k.
\end{align}
Using the notation
\begin{align}\lab{NRCS03}
\d A_k=A_k-\la A_k\ra_{\! f}
=A_k-\7z_k=\d X_k+i\k \d P_k
\end{align}
we have $\la\d A_k\ra_{\! f}=0$ and 
\begin{align*}
0\le \| f \d A_k\|^2&= f\d A_k\d A_k^\dag f^\dag=\la A_k A_k^\dag\ra_{\! f}-|z_k|^2\\
&=\la X_k^2+\k^2P_k^2-i\k [X_k, P_k]\ra_{\! f}-\2x_k^2-\k^2\2p_k^2\\
&=\lp\la X_k^2\ra_{\! f}-\2x_k^2\rp
+\k^2\lp\la P_k^2\ra_{\! f}-\2p_k^2\rp+\hbar \k\\
&=\D_{X_k}^2+\k^2\D_{P_k}^2+\hbar \k 
\end{align*}
where $\D_{X_k}$ and $\D_{P_k}$ are the usual uncertainties of $X_k$ and $P_k$ in the state $f$. Since the quadratic form on the right side must be nonnegative for all real $\k$, its discriminant must be nonpositive, \ie
\begin{align}\lab{NRCS33}
\hbar^2\le 4\D_{X_k}^2\D_{P_k}^2
\end{align}
which is the \sl Heisenberg uncertainty principle. \rm Furthermore, equality holds if and only if $f\d A_k=0$, so $f$ is an eigenvector $\c_\3z$ of $A_k$ with eigenvalue $\7z_k$,
\begin{align}\lab{NRCS4}
\c_\3zA_k=\7z_k \c_\3z.
\end{align}
The $\3x$-representation \eq{NRCS00} of $X_k$ and $P_k$ thus gives
\begin{align}\lab{NRCS4}
x_k \c_\3z(\3x)+\hbar\k\,\pd{\c_\3z(\3x)}{x_k}=\7z_k \c_\3z(\3x)
\end{align}
with a unique normalized solution (up to a constant phase factor)  
\begin{align}\lab{NRCS6}
\c_\3z(\3x)=N'\exp[(\bb z\cdot\3x-\3x^2/2)/\hbar \k]
\end{align}
which requires $\k>0$. Inserting $\bb z=\bt x+i\k\bt p$, \eq{NRCS2} gives
\begin{align}\lab{NRCS7}
\c_\3z(\3x)=N\exp[i\bt p\cdot\3x/\hbar-(\3x-\bt x)^2/2\hbar \k]
\end{align}
with $N=N'\exp(\bt x^2/2\hbar \k)$. These are the Gaussian coherent states in the $\3x$-representation. 

Similarly, in the $\3p$-representation the Fourier transform $\1\c_\3z(\3p)$ satisfies
\begin{align}\lab{NRCS9}
i\hbar\pd{\1\c_\3z(\3p)}{p_k}+i\k p_k\1\c_\3z(\3p)=\7z_k\1\c_\3z(\3p)
\end{align}
giving
\begin{align}\lab{NRCS10}
\1\c_\3z(\3p)&=C'\exp[-i\bb z\cdot \3p/\hbar-\k \3p^2/2\hbar]
=C\exp[-i\bt x\cdot \3p/\hbar-\k(\3p-\bt p)^2/2\hbar]
\end{align}
with $C=C'\exp(\k\bt p^2/2\hbar)$. The physical significance of $\3z=\bt x-i\k\bt y$ is that
\begin{align}\lab{NRCS11}
\bt x=\la \c_\3z |\3X| \c_\3z\ra\qqq \bt p=\la \c_\3z |\3P| \c_\3z\ra,
\end{align}
as required by \eq{NRCS2}. The uncertainties can be read off from the probability densities:
\begin{equation}\lab{DXDP0}\begin{split}
\r(\3x)&\=|\c_\3z(\3x)|^2=N^2\exp[-(\3x-\bt x)^2/\hbar \k]\imp \D_{X_k}=\sr{\hbar\k/2}\\
\2\r(\3p)&\=|\1\c_\3z(\3p)|^2=K^2\exp[-\k(\3p-\bt p)^2/\hbar]\, \imp \D_{P_k}=\sr{\hbar/2\k}
\end{split}\end{equation}
confirming the minimum-uncertainty property
\begin{align}\lab{DXDP1}
\D_{X_k}\D_{P_k}=\hbar/2.
\end{align}

\subsection{The Fundamental Relativistic Quantum States $e_z$}\lab{SS:RCS}

The key to understanding the role of $z$ in quantization is to note that in \eq{phiz00}, $\y\0z$  can be expressed as an inner product
\begin{align}\lab{yzez0}
\y\0z=a e_z^\dag\qq \text{where}\qq e_z\0p=\f_p\0z^*=e^{i\7z\cdot p/\hbar}.
\end{align}
Unlike the plane wave $\f_p\0x$ \eq{PW0} in $M$, $e_z$ is square-integrable with
\begin{align}\lab{SUM3}
\|e_z\|^2=\int_{\O_m}\dd\m\0p\,e^{-2y\cdot p/\hbar}=(\p mc/\l)^\n  K_\n(2\l mc/\hbar)\ \ \ \text{where}\ \  \ \n=\frac{d-1}2
\end{align}
and $K_\n$ is the modified Bessel function.

\rem\lab{R:Bound} \rm 
All wavefunctions $\y\in\5K$ obey the bound
\begin{align}\lab{Bdd0}
|\y\0z|=|a e_z^\dag |\le \| a\| \|e_z\|.\ \clubsuit
\end{align}

The expectations of the \sl Newton-Wigner position operators \rm $X_k$ in $e_z$ at $t=0$ are \ci{K90}
\begin{align}\lab{SUM7}
\la X_k\ra_{e_z}\big |_{t=0}\=\frac{e_z X_k e_z^\dag}{e_z e_z^\dag}
=x_k\,, \qq 1\le k\le d
\end{align}
and the expectations of the energy-momentum operators $P_\m$ in $e_z$ are
\begin{align}\lab{SUM5}
 \la P_\m\ra_{e_z}\=\frac{e_z P_\m e_z^\dag}{e_z e_z^\dag}=m_\l c\,\1y_\m \qq 0\le\m\le d
\end{align}
where
\begin{align}\lab{SUM6}
m_\l=m\,\frac{K_{\n+1}(2\l mc/\hbar)}{K_\n(2\l mc/\hbar)}.
\end{align}

\rem\lab{R:Renorm} \rm 
From the definition 
\begin{align}\lab{Kn00}
K_\n\0z=\int_0^\infty e^{-z\cosh s}\cosh(\n s)\dd s
\end{align}
it follows that
\begin{align}\lab{Kn01}
z>0\imp\pl_\n K_\n\0z=\int_0^\infty e^{-z\cosh s}\sinh(\n s)s\,\dd s>0\ \ \forall \n\ge 0,
\end{align}
hence by \eq{SUM6},
\begin{align}\lab{mlambda00}
m_\l > m\qq \forall \l> 0,
\end{align}
and the \sl effective mass \rm $m_\l$ of the particle in the $(2d+1)$-dimensional \sl state space \rm 
\begin{align}\lab{StateSpace0}
\5T^\l_+=\{x-iy\in\5T_+\colon y^2=\l^2\}
\end{align}
is greater than its `bare' mass $m$. This is a \sl mass renormalization \rm effect due to the convexity of $\O_m$ and the fluctuations of the ray filter $R_y\0p$ \eq{Wy00} around its maximum value at $p_y$.
\rm $\clubsuit$

\prop\lab{NRL} The nonrelativistic limit of $e_z$ at $t=0$ is a Gaussian coherent state. \rm 

\bf Proof: \rm  Setting $c=\hbar=1$,  let
\begin{align*}
u&=\1y=(\sr{1+\3u^2},\3u),\qq \3u=\3y/\l\\
v&=\1p=(\sr{1+\3v^2}, \3v),\qq \3v=\3p/m
\end{align*}
and assume that $y$ is nonrelativistic, \ie $\3u^2\ll 1$. By Proposition \ref{P:scalar}, $e_z\0p$ is negligible unless $p$ is also nonrelativistic, \ie $\3v^2\ll 1$. Then
\begin{align*}
u\app (1+\3u^2/2,\3u)\qqq v\app (1+\3v^2/2,\3v)
\end{align*}
and
\begin{align*}
u\cdot v&\app (1+\3u^2/2)(1+\3v^2/2)-\3u\cdot\3v\app 1+(\3u-\3v)^2/2\\
x\cdot v&=t\sr{1+\3v^2}-\3x\cdot\3v\app t(1+\3v^2/2)-\3x\cdot\3v
\end{align*}
thus
\begin{equation}\lab{ezNR0}\begin{split}
e_z\0p&=e^{-\l m u\cdot v}
\app e^{-\l m}e^{ix\cdot p}e^{-\l m(\3u-\3v)^2/2}\\
&\app e^{-\l m}\exp\LB it(m+\3p^2/2m)-i\3x\cdot\3p-(m\3y-\l\3p)^2/2\RB.
\end{split}\end{equation}
At $t=0$, this is a coherent state with expected position and momentum 
\begin{align}\lab{XPexp0}
\la\3X\ra_z=\3x,\qqq \la \3P\ra_z=m\bh y.
\end{align}
The nonrelativistic free-particle Hamiltonian $H=m+\3p^2/2m$ propagates 
\eq{ezNR0} to time $t$, where it no longer has a minimum uncertainty products. This is not surprising since the uncertainty products is not  Lorentz invariant.  $\blacksquare$

Thus $\5T_+$ can be interpreted as an \sl extended classical phase space \rm for the particle. Since $\dim_\4R\5T_+=2D$ and the classical phase space of a single particle in $d$ space dimensions has dimension $2d=2D-2$, what are the two extra dimensions in $\5T_+$? Clearly, one is the \sl time \rm $t=x_0$. The other is $\l$, which may be called the \sl directivity \rm (Proposition \ref{P:scalar})
or  \sl squeezing parameter \rm of the fundamental states $e_z$ (see Figure \ref{F:Husimi}).

Classical phase spaces are thus submanifolds of $\5T_+$ given by specifying $t$ and $\l$. More generally, choose a  spacelike submanifold of $M$ of codimension one, say
\begin{align}\lab{Sigma00}
\S=\{(x\in M\colon s\0x=0\},
\end{align}
whose normal vector $n_\m\0x=\pl_\m s\0x$ is timelike:\footnote{A more careful analysis \ci{K90} shows that $n$ need only be \sl nowhere spacelike, \rm \ie $n_\m\0x n^\m\0x\ge 0$. We shall not explore this option here.
}
\begin{align}\lab{Sigma01}
n_\m\0x n^\m\0x>0.
\end{align}
In general, what we shall call a \sl covariant classical phase space \rm then has the form
\begin{align}\lab{Pi00}
\G\=\G_{s,\l}=\{x-iy\in \5T_+\colon s\0x=0,\ y^2=\l^2\}=\S-i\O_\l
\end{align}
where $\S$ is a covariant configuration space and
\begin{align}\lab{Olambda0}
\O_\l=\{y\in V_+\colon y^2=\l^2\}
\end{align}
is a relativistic momentum space. To complete the picture, we need a \sl symplectic form \rm on $\G$ which must be Lorentz-invariant in order to give invariant inner products in $\5K$. The cleanest way to do this is to begin with the invariant 2-form \ci{K90}
\begin{align}\lab{alpha00}
\o=\dd x_\m\w \dd y^\m.
\end{align}
A Lorentz-invariant measure on $\G$ is obtained from the $(2d)$-form
\begin{align}\lab{od0}
\o^d \=\o\w \o\w \cdots\w \o 
=d!\,\widehat{\dd x^\m}\w \widehat{\dd y_\m}
\end{align}
where $\widehat{\dd x^\m}\sim\dd x/\dd x_\m$ is a $d$-form with $\dd x_\m$ missing and $\widehat{\dd y_\m}$ is a $d$-form with $\dd y^\m$ missing \ci{K90}.

We have seen that the states $e_z$ generalize the Gaussian coherent states. We call them \sl fundamental relativistic quantum states \rm or simply \sl fundamental states \rm because $e_z$ will be seen to be a natural \sl quantization \rm of $z$ (Remark \ref{R:Quantization}). To see how close they are to being mutually orthogonal, we need the following property from \ci[Section 4.4]{K90}:
\begin{align}\lab{SUM00}
\bx{e_{z} e_{z'}^\dag=(2\p mc/\z)^\n K_\n(mc\z/\hbar),\ \ \n=\frac{d-1}2}
\end{align}
where $K_\n$ is a modified Bessel function of the second kind and
\begin{align}\lab{SUM1}
 \z\=\sr{-w^2}
\end{align}
is the complex length \eq{zeta001} of the complex 
interval\footnote{The conventions in \ci{K90} differ from those used here.  Eqs. \eq{SUM3}, \eq{SUM6} and \eq{SUM00}  reflect the present conventions.
}
\begin{align}\lab{SUM2}
w=z'-\7z=(x'-x)-i(y'+y)\in\5T_+.
\end{align}
Note that when $s\0x=x_0=t$,  then $\dd x_0=0$ and $\o^d/d!$ reduces to the differential form associated with the usual \sl Liouville measure \rm on the classical phase space:
\begin{align}\lab{sx=t}
s\0x=t \imp \o^d/d! \mt \dd^d\3x\w \dd^d\3y.
\end{align}
Hence we define the \sl relativistic Liouville measure \rm as the $(2d)$-form on $\5T_+$ given by
\begin{align}\lab{dg0}
\bx{\dd\g\0z\=N\o^d/ d!}
\end{align}
where the normalization constant $N$ is explained in Proposition \ref{P:IPyz}. Liouville measures covariant with individual phase spaces will be obtained by \sl restricting \rm $\dd\g\0z$ to $\G$.

\defin\lab{D:IPG0} The inner product in $\5K$ with $\G$ as phase space is
\begin{align}\lab{IPy0}
(\y_1\y_2^\dag)_\G=\int_\G\dd\g\0z\, \y_1\0z\y_2\0z^*.
\end{align}
\rm

By the polarization identity, it suffices to work with the norm 
\begin{align}\lab{IPy1}
\|\y\|^2_\G=\int_\G\dd\g\0z\,\r\0z\qq \text{where}\qq \r\0z=|\y\0z|^2.
\end{align}
\rm
 
\prop\lab{P:IPyz} Let $\G$ be a covariant classical phase space of the form \eq{Pi00}. Then for an appropriate choice of $N$ {\rm \ci{K90} we have the `Plancherel theorem'
\begin{align}\lab{IPgen0}
\|\y\|^2_\G=\| a\|^2_{\5H}.
\end{align}
In particular, $\|\y\|^2_\G$ is independent of $\G$ and we may write
\begin{align}\lab{IPgen01}
\|\y\|^2=\int_\G\dd\g\0z\,|\y\0z|^2\qq\forall\ \G.
\end{align}
\rm

This was proved in \ci{K90}, first when $\S$ is a flat time-slice as in \eq{sx=t}, \ie 
\begin{align}\lab{IPgen1}
\|\y\|^2_\G=N\int_{\4R^{2d}}\dd^d\3x\,\dd^d\3y\, \r\0z. 
\end{align}
The integral on the right is of the Liouville type, given the linear relationship \eq{py10} between $\1y$ and the momentum $p_y$.
For general $\G$, we use the fact that the `momentum space' $\O_\l$ is a \sl boundary: \rm
\begin{align}\lab{Olbdy}
\O_\l=-\pl B_\l\ \ \text{where}\ \ B_\l=\{y\in V_+\colon: y^2>\l^2\}
\end{align}
where the minus sign indicates the orientation of $\O_\l$ toward the convex side of the hyperboloid. The contribution from $\l=\infty$ vanishes due to the factor $e^{-y\cdot p/\hbar}$ in the integrand. Stokes' theorem then gives
\begin{align}\lab{OB0}
\int_{\O_\l}\widehat{\dd y_\m}\,\r\0z=-\int_{B_\l}\dd y\,\pd{\r\0z}{y^\m}
\end{align}
and so
\begin{align}\lab{OB1}
\|\y\|^2_\G=-N\int_\S\widehat{\dd x^\m}\int_{B_\l}\dd y\,\pd{\r\0z}{y^\m}.
\end{align}
Choose a world volume $W\subset M$ bounded by two configuration spaces $\S_1, \S_2$ so that
\begin{align}\lab{OB2}
\pl W=\S_2-\S_1
\end{align}
where the corresponding phase spaces (allowing possibly different values of the thermal hyperboloid $\O_\l$) are
\begin{align}\lab{OB3}
\G_1=\S_1-i\O_{\l_1} \qqq \G_2=\S_2-i\O_{\l_2} 
\end{align}
hence
\begin{align}\lab{OB3}
\G_2-\G_1=\pl W-i  \pl \lb B_{\l_2}-B_{\l_1}\rb
=\pl\lb W-i\O_{\l_2}^{\l_1}\rb
\end{align}
where
\begin{equation}\lab{OB4}\begin{split}
\O^{\l_2}_{\l_1}=\pl \lb B_{\l_2}-B_{\l_1}\rb
\end{split}\end{equation}
is a ``thickened'' phase space with $\l_1\le\l \le \l_2$ (assuming 
$\l_1\le\l_2$). Thus $\G_2$ is \sl equivalent \rm to $\G_1$ in the sense that their difference is a boundary:
\begin{align}\lab{OB5}
\G_2=\G_1+\pl Y
\end{align}
where $Y$ is the \sl complex \rm world volume
\begin{align}\lab{OB6}
Y=W-i\O^{\l_2}_{\l_1}\, .
\end{align}

A second application of Stokes' theorem gives
\begin{align}\lab{OB4}
\|\y\|^2_{\G_2}-\|\y\|^2_{\G_1}
=-N\int_W\dd x\int_{B^{\l_2}_{\l_1}}\dd y\,\frac{\pl^2\r\0z}{\pl x_\m\pl y^\m}
\end{align}
where 
\begin{align}\lab{B12}
B^{\l_2}_{\l_1}=\{y\in V_+\colon \l_1^2\le\l^2\le\l_2^2\}.
\end{align}
Since $\Box_z\r\0z=(\Box_z\y)\7\y=-(m/\hbar)^2\r\0z$, it follows that
\begin{equation}\lab{OB5}\begin{split}
\frac{\pl^2\r\0z}{\pl x_\m\pl y^\m}
&=i(\7\pl^\m+\pl^\m)(\7\pl_\m-\pl_\m)\r\0z=i(\7\Box_z-\Box_z)\r\0z=0,
\end{split}\end{equation}
proving that $\|\y\|^2_\G$ is independent of $\G$ as claimed. \ $\blacksquare$

The above proof is not rigorous because it disregards `leaks' that may occur in the integrals \eq{OB4} at spatial infinity. For a rigorous proof, see \ci{K90}.

\defin\lab{D:jJ} The {\bf microlocal current} $j_\m\0z$ in $\5T_+$ and the {\bf local current} $J_\m\0x$ in $M$ generated by $\r\0z=|\y\0z|^2$ are given by
\begin{equation}\lab{jJ00}\begin{split}
\bx{j_\m\0z=-N\,\pd{\r\0z}{y^\m}\qqq J_\m\0x\=\int_{B_\l}\dd y\,j_\m(x-iy).}
\end{split}\end{equation}
In terms of $\y\0z$,
\begin{align}\lab{jJ01}
j_\m\0z=-2N\im (\pl_\m\y\0z\!\cdot\!\y\0z^*)
=iN\y\0z\stack{\lra}{\pl_\m}\y\0z^*
\end{align}
where
\begin{align}\lab{jJ02}
\y_1\stack{\lra}{\pl_\m}\y_2^*\=(\pl_\m\y_1)\y_2^*-\y_1\pl_\m\y_2^*.
\end{align}
\rm

By \eq{OB5}, both currents are conserved in $M$, \ie with respect to variations of $\S$:
\begin{align}\lab{OB7}
\pd{j_\m(x-iy)}{x_\m}=\pd{J_\m\0x}{x_\m}=0.
\end{align}
This makes
\begin{align}\lab{OB7}
\|\y\|^2=\int_\S\widehat{\dd x^\m}\,J_\m\0x
=\int_\S\widehat{\dd x^\m}\int_{B_\l}\dd y \,j_\m(x-iy)
\end{align}
the \sl total charge \rm of the conserved current $J_\m\0x$ over $\S$, as well as that of $j_\m\0z$ over $\S\times B_\l$. Being conserved, $\|\y\|^2$ is independent of $\G$, so we can drop the subscript in $\|\y\|^2_\G$\,. 

\rem\lab{R:RelBorn} \rm The fact that the norm
\begin{align}\lab{IPy1}
\bx{\|\y\|^2=\int_\G\dd\g\0z\,|\y\0z|^2=\|a\|^2_{\5H}}
\end{align}
is independent of $\G$ proves that $|\y\0z|^2$ is a valid  probability density for \sl  every \rm covariant phase space. This is a  
\sl Lorentz-invariant  version of the nonrelativistic Born Rule. \rm
\ $\clubsuit$

\rem\lab{R:RU} Resolution of Unity in terms of the fundamental states:  \rm \HB
Expressing \eq{IPy1} as
\begin{align}\lab{RU00}
\y\y^\dag=\int_\G\dd\g\0z\,\y e_z^\dag e_z\y^\dag
\end{align}
and peeling away the factors $\y$ and $\y^\dag$ gives the operator equation
\begin{align}\lab{RU01}
\bx{\int_\G\dd\g\0z\, e_z^\dag e_z=I}
\end{align}
for all $\G$ of the form \eq{Pi00}, where $I$ is the identity operator on $\5H$. This is a relativistic version of the standard resolution of unity in terms of Gaussian coherent states. \rm $\clubsuit$

\rem\lab{R:Reprod} \rm By the Resolution of Unity \eq{RU01},
\begin{align}\lab{RKHS}
\bx{\y(z')=a e_{z'}^\dag=\int_\G\dd\g\0z \,a e_z^\dag e_z e_{z'}^\dag
=\int_\G\dd\g\0z \,\y\0z K(z'-\7z).}
\end{align}
If $z'\in\G$, this is a \sl reproducing property \rm generalizing that of the Gaussian coherent states. If $z'\notin\G$, then $z'$ is either in the future ($z'>\G$) or past ($z'<\G$) of $\G$ and \eq{RKHS} \sl propagates \rm $\y$ from $\G$ to $z'$. Thus \sl $K$ unifies the ideas of reproducing kernel and  propagator \rm in $\5K$. Figure \ref{F:Husimi} shows the behavior of  
$K(z'-\7z)$, which measures the \sl correlation \rm between $e_{z'}$ and $e_z$. \rm $\clubsuit$

\begin{figure}[t]\begin{center}
\includegraphics[width=4 in]{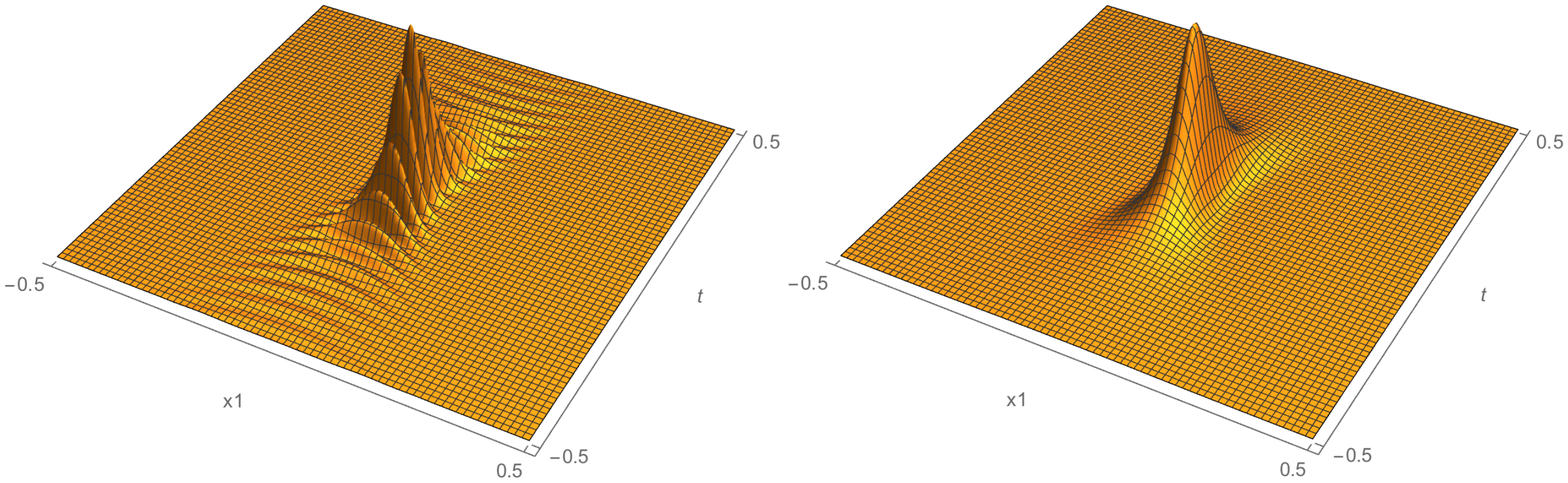}\\ \sv{-28}
\includegraphics[width=4 in]{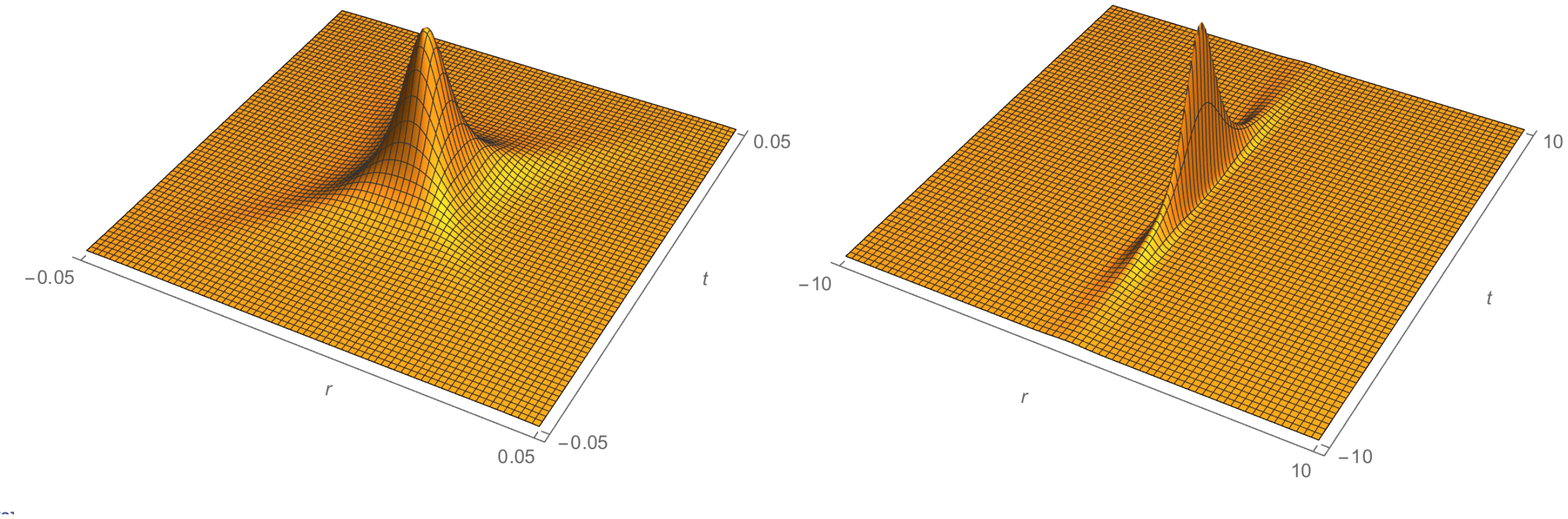} \sv{-15}
\caption{Plots of the reproducing/propagating kernel $K$ \eq{RKHS}. \HB
\sl Top left: \rm Plot of $(\!\re K\0w)^2$ with $d=3$ and $w=z'-\7z=(t-i\l, x_1,0,0)$ \HB
 \sl Top right: \rm Plot of $|K\0w|^2$ with the same parameters  \HB
 \sl Bottom left: \rm Plot of $|K\0w|^2$ with $\l mc=0.1\hbar$ \HB
\sl Bottom right: \rm Plot of $|K\0w|^2$ with 
$\l mc=20\hbar$. \HB
The quadrature complement $(\!\im K)^2$ of $(\!\re K)^2$ has similar oscillations with offset phases, making the sum $|K|^2$ smooth. The level surfaces $\5B_\s$ and $\5W_\t$  \eq{BW00} give the shape of the beam and its wave fronts, respectively, and $\l m$ controls the \sl directivity \rm of the beam.} \HB \label{F:Husimi} 
\end{center}\end{figure}
 
\rem\lab{R:Quantization}  Quantization of $\5T_+$. \rm 
The mapping 
\begin{align}\lab{Q00}
Q\colon \5T_+\to \5H,\qq z\mt e_z
\end{align}
sends the (extended) \sl classical \rm state $z$ to the \sl quantum \rm state $e_z$, so it may be viewed as a `quantization' of $\5T_+$.  $\clubsuit$

\rem\lab{R:Dirac2} \rm 
Fundamental fundamental states for Dirac particles  are defined in \ci{K90}. \rm $\clubsuit$

\subsection{Bohmian Mechanics of a Single Relativistic Particle}\lab{SS:Thermal}

The density matrix of a quantum-mechanical canonical ensemble, representing a system in thermal equilibrium with a heat bath at absolute temperature $T$, is\footnote{We take Boltzmann's constant $k_B=1$, so that $T$ has units of \sl energy. \rm
}
\begin{align}\lab{bH0}
\r=Z\inv e^{-\b H},\qq \b=T\inv
\end{align}
where $H$ is the system's Hamiltonian operator and
\begin{align}\lab{bH1}
Z=\tr e^{-\b H}
\end{align}
is the partition function. 
The \sl thermal expectation value  \rm of an operator $A$ is
\begin{align}\lab{bH2}
\la A\ra=\Tr(A\r)=Z\inv\Tr(A e^{-\b H}).
\end{align}
Formally, it is possible to build a statistical thermodynamics of a single relativistic particle. For the massive scalar under consideration, the main thermodynamic potentials of the probability distribution $\r\0z$ are: 

\bul  \sl Internal energy \rm $U\=\la H \ra=-Z\inv\pl_\b Z=-\pl_\b\ln Z$.

\bul  \sl Entropy \rm $S\=-\la\ln\r\ra=-\Tr(\r\ln\r)=\F+\b U$
where $\F=\ln Z$ is the \sl Massieu potential.

\bul \sl Helmholtz free energy \rm $F=U-TS=-\F/\b$.

This begs the question: {\sl\blue What is the \bf classical ensemble \sl leading to the above potentials?}

\rem\lab{R:Trace} \rm The trace of an operator can be computed in the $z$-representation by
\begin{align}\lab{Trace0}
\Tr B=\int_\G\dd\g\0z\,\2B\0z\ \ \text{where}\ \ 
\2B\0z=e_z B e_z^\dag.
\end{align}
As a partial (and far from rigorous) proof, consider the rank 1 operator $e_w^\dag e_w$. By \eq{RU01},
\begin{align}\lab{Trace1}
\int_\G\dd\g\0z\,e_z e_w^\dag e_w e_z^\dag
=\int_\G\dd\g\0z\,e_w e_z^\dag e_z e_w^\dag=e_w e_w^\dag
=\Tr (e_w^\dag e_w ),
\end{align}
confirming \eq{Trace0}. \rm $\clubsuit$

Applying \eq{Trace0} to \eq{bH2} gives
\begin{equation}\lab{bH03}\begin{split}
\Tr(A e^{-\b H})=\Tr(e^{-\b H/2}Ae^{-\b H/2})
=\int_\G\dd\g\0z e_z e^{-\b H/2}Ae^{-\b H/2}e_z^\dag.
\end{split}\end{equation}
To keep this covariant, choose a unit vector $u\in V_+$ representing a possible time axis $\4R u$ and let
\begin{align}\lab{bH5}
H=u\cdot p,
\end{align}
so that $u=(1,\30)$ gives the usual energy. By \eq{yzez0},
\begin{align}\lab{bH6}
e_z e^{-\b H/2}\0p
=e^{(i\7z-\vq)\cdot p/\hbar}=e_{z-i\vq}\0p
\end{align}
where
\begin{align}\lab{bH10}
\vq\=\tfrac12 \hbar \b u.
\end{align}
Since its magnitude $|\vq|=\tfrac12\hbar\b$ gives the  ``equilibrium temperature'' and its spacetime direction $u=\vq/|\vq|$ gives the ``equilibrium frame,'' we call $\vq$ the \bf thermal vector. \rm

\thm\lab{T:zthermal}
The thermal expectation $\la A\ra$ can be expressed entirely in terms of integrals over $z\in\G$ for any phase space $\G=\S-i\O_\l\subset\5T_+$ by
\begin{align}\lab{bH11}
\bx{\la A\ra\0\vq=\2Z\0\vq\inv\int_\G\dd\g\0z \, \2A(z-i\vq)}
\end{align}
where
\begin{align}\lab{bH010}
\2A\0z=e_z A e_z^\dag\ \ \ \text{\rm and}\ \ \ 
\2Z\0\vq=\int_\G\dd\g\0z \, \Vert e_{z-i\vq}\Vert^2.
\end{align}
\rm

Equation \eq{bH11} can be expressed in the suggestive form
\begin{align}\lab{bH12}
\la A\ra\0\vq=\frac{\int_\G\dd\g\0z \,  e_{z-i\vq} A  e_{z-i\vq}^\dag}
{\int_\G\dd\g\0z \,  e_{z-i\vq} e_{z-i\vq}^\dag}.
\end{align}

\bf Proof: \rm  Assuming the validity of \eq{Trace0},
\eq{bH11} follows from \eq{bH03} and \eq{bH6}; $\2Z\0\vq$ is a special case with $A=I$.  \ $\blacksquare$

Note that the \sl \sl thermal translation  \rm
\begin{align}\lab{bH9}
z\mt z-i\vq=x-i(y+\vq)
\end{align}
is \sl internal: \rm it leaves $x$ invariant while dragging
 $y$ further away from the origin since
\begin{align}\lab{y+vq}
|y+\vq|=\sr{\lp |y|^2+|\vq|^2+2y\cdot\vq\rp}\ge |y|+|\vq|
\end{align}
by \eq{RTI0}.

For given $\vq\ne 0$ (\ie $T<\infty)$, \eq{bH9} breaks Lorentz symmetry as it selects a preferred equilibrium frame through $u\in V_+$. That symmetry is restored if we allow $\5G_0$ to act on the set thermal expectations in all equilibrium frames by 
\begin{align}\lab{bH09}
\la A\ra U(\L)\0\vq= \la A\ra(\vq\L).
\end{align}

\rem\lab{R:Ensemble}\rm  Theorem \ref{T:zthermal} answers the question posed above: \rm 

{\blue 
Every phase-space element $z\in\G$ in \eq{bH11} represents a unique \sl classical phase space trajectory \rm of the particle in $\5T_+$, and the integral $\int_\G\dd\g\0z$ sums over all such trajectories. The ensemble average is independent of the phase space $\G$ since each $\G$ intersects every classical trajectory exactly once and $\dd\g\0z$ is $\5G_0$-invariant. This  proves that $\la A\ra$ depends on the \sl trajectories  \rm and not their individual points. {\bf These trajectories are the microstates of our ensemble. }
}

I hold Remark \ref{R:Ensemble} to be the main result of this investigation. Together with holomorphic gauge theory, which proposes a way to introduce interactions without destroying holomorphy,  it gives a solid foundation to relativistic Bohmian mechanics.

\rem\lab{R:RMB2} \rm 
For the free particle considered here, the classical trajectories are straight lines. In the next section we propose to include interactions with a gauge field by postulating a \sl fiber metric \rm in the quantum Hilbert space. At the quantum level, the fiber metric subjects the wave function to a gauge interaction. At the classical level, it distorts the trajectories of the associated classical particle to reflect that interaction.\  $\clubsuit$

\section{Interactions via Holomorphic Gauge Theory}\lab{S:HGT}
So far we have dealt exclusively with a single free relativistic particle. The requirement that  wave functions be holomorphic makes it difficult to introduce interactions through potentials as done in the nonrelativistic theory. We shall instead introduce them \sl covariantly \rm through a method we call  \sl Holomorphic Gauge Theory \rm \ci{K81}. 
The probability density $\r\0z=|\y\0z|^2$  and the microlocal current $j_\m\0z$ \eq{jJ00} are invariant under  global  gauge transformations $\y\0z\mt \y\0z\c$, where $\c$ is constant with $|\c|\=1$. But they are not invariant under \sl microlocal \rm gauge transformations, where $\c\0z$ is holomorphic in $\5T_+$ to preserve the holomorphy of $\y$, since $|\c\0z|\=1$  implies that $\c$ is constant. To admit microlocal gauge transformations, we introduce a  \sl fiber metric\rm\footnote{Unless $g$ is constant, it cannot be holomorphic. Still, we write $g\0z$ instead of $g(z,\7z)$ for brevity.
}
$g\0z>0$ into the Hilbert space $\5K$, so that the norm
\eq{IPy1} becomes
\begin{align}\lab{HGT2}
\|\y\|^2_\G=\int_\G\dd\g\0z \,\r\0z
\end{align}
where
\begin{align}\lab{HGT02}
\r\0z=\y\0z g\0z\y\0z^*
\end{align}
is to be interpreted as the particle's \sl  covariant probability  density. \rm  Then $\r\0z$ is invariant under $\y'\0z = \y\0z \c\0z$ if and only if $g\0z$ absorbs the factor $\c\0z$ and its conjugate:
\begin{align}\lab{HGT3}
\r '\0z=\r \0z \ \Iff \   g'\0z =\c\0z\inv g\0z (\c\0z^*)\inv.
\end{align}
To find the gauge potential and its curvature field, use the exterior derivative
\begin{align}\lab{HGT8}
\dd=\dd x^\m\pd{}{x^\m}+\dd y^\m\pd{}{y^\m}
=\dd z^\m\pd{}{z^\m}+\dd \7z^\m\pd{}{\7z^\m}=\pl+\7\pl
\end{align}
where $\pl$ and $\7\pl$ are the holomorphic and antiholomorphic exterior derivatives, with
\begin{align}\lab{d7d}
\dd^2=0 \ \Iff \ \pl^2=\7\pl^2=\pl\7\pl+\7\pl\pl=0.
\end{align}
Since $\pl\y\0z^*=0$, 
\begin{align}\lab{HG0}
\pl\r=\pl(\y g\y^*)&=\pl(\y g)\y^*
=(\pl\y+\y\pl g\cdot g\inv)g\y^*
=(\5D\y)g\y^*
\end{align}
where $\5D$ is the \sl holomorphic exterior derivative \rm
\begin{align}\lab{HGT11}
\5D\y\0z=\pl(\y g)g\inv=\pl\y+\y\pl g\cdot g\inv\equiv\pl\y+\y\5A
\end{align}
with the potential 1-form
\begin{align}\lab{HGT12}
\5A=\pl g\cdot g\inv=\pl\ln g.
\end{align}
In the abelian case, the \sl gauge field \rm is given by
\begin{align}\lab{HGT13}
\5F\0z&=\dd \5A\0z=\7\pl\5A=\7\pl\pl\ln g\0z.
\end{align}
We have thus arrived at a form of the electromagnetic field as a  holomorphic gauge theory for a massive scalar. The potential form $\5A$ is related to the electromagnetic $D$-potential 1-form $A$ by
\begin{align}\lab{HGT013}
\5A\0z=iA\0z=(\pl_\m \ln g\0z)\dd z^\m.
\end{align}
Thus $\5A\colon \5T_+\to\4C$ is derived from a \sl  superpotential \rm  $\ln g\0z$, something impossible in $M$.

The same conservation law \eq{OB5} making the free norm \eq{IPy1}   invariant can now be applied to \eq{HGT02}. We find
\begin{equation}\lab{HGT4}\begin{split}
\frac{\pl^2\r \0z}{\pl x_\m\pl y^\m}
&=i(\7\pl^\m+\pl^\m)(\7\pl_\m-\pl_\m)\r \0z
=i(\7\Box_z-\Box_z)(\y g\y^\dag)\\
&=i\y\7\Box_z(g\y^\dag)-i\lb\Box_z (\y g)\rb\y^\dag\\
&=i\y[\Box_z(\y g)]^\dag -i\lb\Box_z (\y g)\rb\y^\dag.
\end{split}\end{equation}
A necessary and sufficient condition for conservation of probability is therefore
\begin{align}\lab{HGT5}
-\Box_z (\y\0zg\0z)=\y\0z M\0z,\ \ \text{where}\ \  M\0z=M\0z^*
\end{align}
replaces the factor $(mc/\hbar)^2$ in \eq{HKG0}. $M\0z$ thus plays the role of a \sl mass-squared operator with the gauge-field interactions built in covariantly.  \rm 

For particles with internal symmetry, the above scalar gauge theory extends to a \sl non-abelian \rm gauge theory where the fiber metric $g\0z$ is a Hermitian $n\times n$ matrix and the gauge potential is given by the matrix-valued 1-form
\begin{align}\lab{HYM0}
\5A=\pl g\cdot g\inv.
\end{align}
Since $\pl g$ need not commute with $g\inv$,
 $\5A$ cannot generally be expressed in the form $\pl\ln g$. Hence the non-abelian gauge potential cannot be derived from a superpotential.

The gauge field is given by\footnote{The usual expression for the curvature on a non-abelian gauge field is $\5F=\7\pl\5A+\pl\5A+\5A\w\5A$. The sign difference is due to the fact that exterior derivatives act to the \sl right \rm while our operators act to the \sl left. \rm
}
\begin{align}\lab{HYM2}
\5F=\dd\5A-\5A\w\5A=\7\pl\5A+\pl\5A-\5A\w\5A.
\end{align}
However, \eq{HYM0} implies the \sl integrability condition \rm
\begin{align}\lab{HYM1}
\pl\5A=-\pl g\w\pl g\inv=\pl g\w(g\inv \pl g\cdot g\inv)=\5A\w\5A,
\end{align}
giving
\begin{align}\lab{HYM2}
\5F=\7\pl\,\5A.
\end{align}
The integrability condition thus extends the linear relation between  potential and field from the scalar case to the non-abelian case. This -- and the superpotential $\ln g$ \eq{HGT12} in the scalar case -- gives \sl holomorphic \rm gauge theory a status not shared by ordinary gauge theory.

\rem\lab{R:Grav} \rm 
Holomorphic gauge theory brings gauge theory closer to General Relativity. The former uses a metric defined on the Hilbert space of quantum states while the latter uses a metric on tangent spaces. In the present context, this metric would have the form $g_{\m\n}(z,z^*)$ with $z$ and $z^*$ formally independent, representing a map
\begin{align}\lab{gmn0}
g_{\m\n}\colon \5T_+\times\5T_+^*\to\4C.
\end{align}
Just as the Einstein metric is expected to distort free-particle trajectories to follow gravity, so is the fiber metric $g\0z$ expected to distort them to follow the gauge field $\5F$. This will be the subject of future work. \  $\clubsuit$

\section*{Acknowledgement}
I thank Sheldon Goldstein for inspiring me to study Bohmian Mechanics.

\VE


\begin{thebibliography}{00}

		
		
\bibitem{DGZ92} D Dürr, S Goldstein, and N Zanghì, Quantum equilibrium and the origin of absolute uncertainty. \sl J Stat Phys \rm\textbf{67}, 843--907 (1992).
\url{https://doi.org/10.1007/BF01049004}

\bibitem{GJ87} J Glimm and A Jaffe, \sl Quantum Physics: A Functional Integral Point of View, \rm 2nd ed.
Springer, 1987

\bibitem{H93} P R Holland, \sl The Quantum Theory of Motion. \rm Cambridge University Press, 1993


\bibitem{K81} G Kaiser, Phase-space approach to relativistic quantum mechanics. III. Quantization, relativity, localization and gauge freedom.
\sl J. Math. Phys. \bf 22\rm, 705 (1981).  
\url{https://doi.org/10.1063/1.524962}

\bibitem{K90} G Kaiser,  \sl Quantum Physics, Relativity, and Complex Spacetime: Towards a New Synthesis. \rm North Holland, 1990. \url{https://arxiv.org/abs/0910.0352}

\bibitem{K00} G Kaiser, Complex-distance potential theory and hyperbolic equations, in \sl Clifford Analysis, \rm J. Ryan and W. 
Spr\"ossig, eds., Birkh\"auser, Boston, 2000. \url{https://arxiv.org/abs/math-ph/9908031}

\bibitem{SW64} R F Streater and A S Wightman, \sl PCT, Spin and Statistics, and All That. \rm Princeton University Press, 2001

\bibitem{K11} G Kaiser, \sl A Friendly Guide to Wavelets. \rm
 Modern Birkh\"auser Classics, Boston, 2011

			
\end{thebibliography}
\end{document}